\PassOptionsToPackage{unicode}{hyperref}
\PassOptionsToPackage{hyphens}{url}
\PassOptionsToPackage{dvipsnames,svgnames,x11names}{xcolor}
\documentclass[
]{article}

\usepackage{amsmath,amssymb}
\usepackage{iftex}
\ifPDFTeX
  \usepackage[T1]{fontenc}
  \usepackage[utf8]{inputenc}
  \usepackage{textcomp} % provide euro and other symbols
\else % if luatex or xetex
  \usepackage{unicode-math}
  \defaultfontfeatures{Scale=MatchLowercase}
  \defaultfontfeatures[\rmfamily]{Ligatures=TeX,Scale=1}
\fi
\usepackage{lmodern}
\ifPDFTeX\else  
\fi
\IfFileExists{upquote.sty}{\usepackage{upquote}}{}
\IfFileExists{microtype.sty}{% use microtype if available
  \usepackage[]{microtype}
  \UseMicrotypeSet[protrusion]{basicmath} % disable protrusion for tt fonts
}{}
\makeatletter
\@ifundefined{KOMAClassName}{% if non-KOMA class
  \IfFileExists{parskip.sty}{%
    \usepackage{parskip}
  }{% else
    \setlength{\parindent}{0pt}
    \setlength{\parskip}{6pt plus 2pt minus 1pt}}
}{% if KOMA class
  \KOMAoptions{parskip=half}}
\makeatother
\usepackage{xcolor}
\makeatletter
\ifx\paragraph\undefined\else
  \let\oldparagraph\paragraph
  \renewcommand{\paragraph}{
    \@ifstar
      \xxxParagraphStar
      \xxxParagraphNoStar
  }
  \newcommand{\xxxParagraphStar}[1]{\oldparagraph*{#1}\mbox{}}
  \newcommand{\xxxParagraphNoStar}[1]{\oldparagraph{#1}\mbox{}}
\fi
\ifx\subparagraph\undefined\else
  \let\oldsubparagraph\subparagraph
  \renewcommand{\subparagraph}{
    \@ifstar
      \xxxSubParagraphStar
      \xxxSubParagraphNoStar
  }
  \newcommand{\xxxSubParagraphStar}[1]{\oldsubparagraph*{#1}\mbox{}}
  \newcommand{\xxxSubParagraphNoStar}[1]{\oldsubparagraph{#1}\mbox{}}
\fi
\makeatother

\providecommand{\tightlist}{%
  \setlength{\itemsep}{0pt}\setlength{\parskip}{0pt}}\usepackage{longtable,booktabs,array}
\usepackage{calc} % for calculating minipage widths
\usepackage{etoolbox}
\makeatletter
\patchcmd\longtable{\par}{\if@noskipsec\mbox{}\fi\par}{}{}
\makeatother
\IfFileExists{footnotehyper.sty}{\usepackage{footnotehyper}}{\usepackage{footnote}}
\makesavenoteenv{longtable}
\usepackage{graphicx}
\makeatletter
\def\maxwidth{\ifdim\Gin@nat@width>\linewidth\linewidth\else\Gin@nat@width\fi}
\def\maxheight{\ifdim\Gin@nat@height>\textheight\textheight\else\Gin@nat@height\fi}
\makeatother
\setkeys{Gin}{width=\maxwidth,height=\maxheight,keepaspectratio}
\makeatletter
\def\fps@figure{htbp}
\makeatother
\NewDocumentCommand\citeproctext{}{}

\makeatletter
 \let\@cite@ofmt\@firstofone
 \def\@biblabel#1{}
 \def\@cite#1#2{{#1\if@tempswa , #2\fi}}
\makeatother
\newlength{\cslhangindent}
\newlength{\csllabelwidth}
\newenvironment{CSLReferences}[2] % #1 hanging-indent, #2 entry-spacing
 {\begin{list}{}{%
  \setlength{\itemindent}{0pt}
  \setlength{\leftmargin}{0pt}
  \setlength{\parsep}{0pt}
  \ifodd #1
   \setlength{\leftmargin}{\cslhangindent}
   \setlength{\itemindent}{-1\cslhangindent}
  \fi
  \setlength{\itemsep}{#2\baselineskip}}}
 {\end{list}}
\usepackage{calc}

\makeatletter
\@ifpackageloaded{caption}{}{\usepackage{caption}}
\AtBeginDocument{%
\ifdefined\contentsname
  \renewcommand*\contentsname{Table of contents}
\else
  \newcommand\contentsname{Table of contents}
\fi
\ifdefined\listfigurename
  \renewcommand*\listfigurename{List of Figures}
\else
  \newcommand\listfigurename{List of Figures}
\fi
\ifdefined\listtablename
  \renewcommand*\listtablename{List of Tables}
\else
  \newcommand\listtablename{List of Tables}
\fi
\ifdefined\figurename
  \renewcommand*\figurename{Figure}
\else
  \newcommand\figurename{Figure}
\fi
\ifdefined\tablename
  \renewcommand*\tablename{Table}
\else
  \newcommand\tablename{Table}
\fi
}
\@ifpackageloaded{float}{}{\usepackage{float}}
\floatstyle{ruled}
\@ifundefined{c@chapter}{\newfloat{codelisting}{h}{lop}}{\newfloat{codelisting}{h}{lop}[chapter]}
\floatname{codelisting}{Listing}

\makeatother
\makeatletter
\@ifpackageloaded{caption}{}{\usepackage{caption}}
\@ifpackageloaded{subcaption}{}{\usepackage{subcaption}}
\makeatother

\ifLuaTeX
  \usepackage{selnolig}  % disable illegal ligatures
\fi

\usepackage{pos}
\usepackage{bookmark}

\IfFileExists{xurl.sty}{\usepackage{xurl}}{} % add URL line breaks if available
\hypersetup{
  pdftitle={Truncation orders, external constraints, and the determination of \textbar V\_\{cb\}\textbar{}},
  colorlinks=true,
  linkcolor={blue},
  filecolor={Maroon},
  citecolor={Blue},
  urlcolor={Blue},
  pdfcreator={LaTeX via pandoc}}

\title{Truncation orders, external constraints, and the determination of
\(|V_{cb}|\)}

\author*[a]{Eric Persson}
\author[a]{Florian Bernlochner}

\affiliation[a]{Physikalisches Institut, Bonn University,\\
  Bonn, Germany}

\abstract{We present a model selection framework for the extraction of
the CKM matrix element \(|V_{cb}|\) from exclusive \(B \to D^* l \nu\)
decays. By framing the truncation of the Boyd-Grinstein-Lebed (BGL)
parameterization as a model selection task, we apply the Akaike
Information Criterion (AIC) to choose the optimal truncation order. We
demonstrate the performance of our approach through a comprehensive toy
study, comparing it to the Nested Hypothesis Test (NHT) method used in
previous analyses. Our results show that the AIC-based approach produces
unbiased estimates of \(|V_{cb}|\), albeit with some issues of
undercoverage. We further investigate the impact of unitarity
constraints and explore model averaging using the Global AIC (gAIC)
approach, which produced unbiased results with correct coverage
properties. Our findings suggest that model selection techniques based
on information criteria and model averaging offer a promising path
towards more reliable \(|V_{cb}|\) determinations.}

\FullConference{
  42nd International Conference on High Energy Physics (ICHEP) \\
  18-24 July, 2024 \\
  Prague, Czechia
}

\begin{document}
\maketitle

\section{Introduction}\label{introduction}

The precise determination of the Cabibbo-Kobayashi-Maskawa (CKM) matrix
element \(|V_{cb}|\) remains a crucial challenge in flavor physics. A
long-standing tension exists between the values obtained from inclusive
and exclusive measurements of semileptonic B meson decays. This
discrepancy, often referred to as the \(|V_{cb}|\) puzzle, has
significant implications for our understanding of quark flavor physics
and potential new physics beyond the Standard Model.

In recent years, much attention has focused on the exclusive decay
channel \(B \to D^* l \nu\), which offers a promising avenue for precise
\(|V_{cb}|\) determination due to its experimental accessibility and
theoretical cleanliness (Bordone and Juttner 2024). However, the
extraction of \(|V_{cb}|\) from this channel is sensitive to the
parameterization of hadronic form factors, which describe the strong
interaction effects in the decay.

The results presented in this paper are preliminary findings from a more
comprehensive study (F. Bernlochner et al. forthcoming).

\subsection{BGL Parameterization and the Truncation
Dilemma}\label{bgl-parameterization-and-the-truncation-dilemma}

The decay rate in the \(B \to D^* l \nu\) channel can be expressed in
terms of three form factors for massless leptons. These form factors can
be parameterized using the Boyd-Grinstein-Lebed (BGL) expansion (Boyd,
Grinstein, and Lebed 1995), which exploits the analytic properties of
the form factors and incorporates constraints from unitarity. The BGL
parameterization expresses each form factor as an infinite series:

\begin{equation}\phantomsection\label{eq-form-factor}{ 
f(z) = \frac{1}{P(z)\phi(z)} \sum_{n=0}^{\infty} a_n z^n
}\end{equation}

where \(z\) is a kinematic variable, \(P(z)\) is a Blaschke factor,
\(\phi(z)\) is an outer function, and \(a_n\) are the expansion
coefficients (Simons, Gustafson, and Meurice 2024). Similar expansions
with coefficients \(b_n\) and \(c_n\) are used for the other two form
factors.

In practice, these infinite series must be truncated at some finite
order, represented by three integers \((N_a, N_b, N_c)\). This
truncation presents a fundamental dilemma: truncating too early may
introduce bias in our estimates of \(|V_{cb}|\), while truncating too
late leads to an overly complex model, unnecessarily increasing the
variance of the fit results. This situation exemplifies the classic
bias-variance trade-off in statistical modeling. The choice of
truncation order can significantly impact the extracted value of
\(|V_{cb}|\), as demonstrated in previous studies (F. U. Bernlochner,
Ligeti, and Robinson 2019; Gambino, Jung, and Schacht 2019).

\section{Model Selection Framework}\label{model-selection-framework}

In the present paper we propose framing this truncation problem as a
model selection task, drawing on the well-established field of
statistical model selection. In this framework, each possible BGL
truncation order \((N_a, N_b, N_c)\) represents a distinct model. The
goal is to choose both the best model (truncation order) and the best
parameters within that model.

By framing our problem in these terms, we can apply established
statistical techniques and provide a principled, rigorous procedure for
choosing the truncation order. This approach aims to minimize arbitrary
choices on the part of the researcher, enhancing the reproducibility and
reliability of \(|V_{cb}|\) determinations.

\subsection{Components of Model
Selection}\label{components-of-model-selection}

To clarify the choices involved in model selection, we propose the
following taxonomy:

\begin{enumerate}
\def\labelenumi{\arabic{enumi}.}
\item
  \textbf{Model evaluation metrics}: These assess model performance
  quantitatively. Common examples include the sum of squared errors,
  chi-square statistic, Akaike Information Criterion (AIC), Bayesian
  Information Criterion (BIC), and cross-validation error.
\item
  \textbf{Model selection decision rules}: These determine how to choose
  between competing models based on their evaluation metrics. For
  instance, one might select the model with the lowest AIC or choose a
  more complex model only if it improves the metric by a certain
  threshold.
\item
  \textbf{Model space search algorithms}: These define how to navigate
  through the space of possible models. Examples include forward
  stepwise selection, backward elimination, and exhaustive search.
\end{enumerate}

Previous work in this field can be analyzed using this taxonomy.
Bernlochner et al. (F. U. Bernlochner, Ligeti, and Robinson 2019) used
sum of squared residuals (SSR) as their metric, chose a more complex
model only if the improvement in SSR exceeded 1, and employed forward
stepwise selection as their search algorithm. Gambino et al. (Gambino,
Jung, and Schacht 2019) adopted a similar approach, but imposed
unitarity constraints and moved towards more complex models until
\(|V_{cb}|\) estimates stabilized.

In contrast, we propose using the Akaike Information Criterion (AIC)
(Akaike 1974) as our metric, choosing the model with the lowest AIC, and
employing an exhaustive search within feasible models. We argue that
these choices are more less arbitrary, simpler to implement, and easier
to interpret statistically than previous approaches.

\subsection{Akaike Information Criterion
(AIC)}\label{akaike-information-criterion-aic}

The Akaike Information Criterion (AIC) is a well-established model
selection tool in statistics that provides a principled way to balance
model complexity against goodness-of-fit.

The AIC is defined as:

\begin{equation}\phantomsection\label{eq-aic}{
\text{AIC} = -2 \log(L) + 2k
}\end{equation}

where \(L\) is the maximum likelihood of the model and \(k\) is the
number of parameters.

Despite its apparent simplicity, the AIC has strong theoretical
motivations (Blankenshipa, Perkinsb, and Johnsonc 2002; Burnham and
Anderson 1998). It has deep connections to information theory,
specifically to the Kullback-Leibler divergence between the true
data-generating process and the model. It provides a principled way to
balance model complexity against goodness-of-fit, addressing the
bias-variance trade-off. Furthermore, it is widely applicable across
various fields and model types, making it a versatile tool for model
selection.

In the context of BGL parameterization, the AIC offers a natural way to
determine the optimal truncation order, potentially leading to more
robust and reliable \(|V_{cb}|\) extractions.

\section{Toy Study Results}\label{toy-study-results}

To evaluate our AIC-based approach against the Nested Hypothesis Test
(NHT) method used in (F. U. Bernlochner, Ligeti, and Robinson 2019), we
conducted a comprehensive toy study. We simulated \(B \to D^* l \nu\)
decay data based on realistic parameters, assumed an underlying true BGL
order to generate the data, and used the Belle covariance matrix (Heavy
Flavor Averaging Group 2024) to incorporate a realistic error structure.
We then fitted models using both the NHT and AIC procedures. Our primary
goal was to demonstrate that our method produces unbiased estimates of
\(|V_{cb}|\) with correct coverage properties.

We present our results primarily through ``pull distributions'', defined
as:

\begin{equation}\phantomsection\label{eq-pull}{
\text{pull} = \frac{\text{estimated value} - \text{true value}}{\text{estimated uncertainty}}.
}\end{equation}

\begin{figure}

\centering{

\includegraphics[width=0.75\textwidth,height=\textheight]{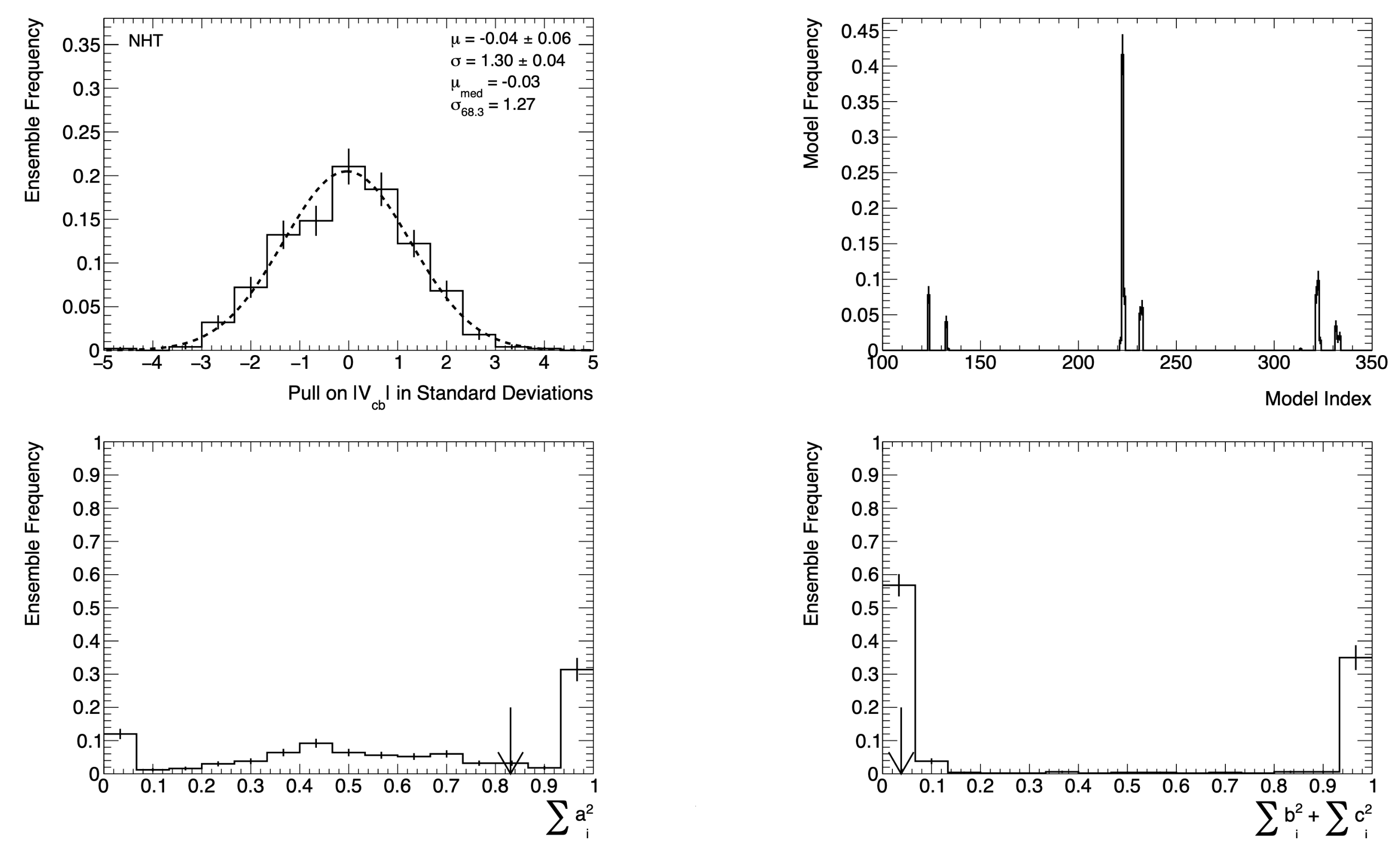}

}

\caption{\label{fig-nht-no-ut}Pull plot of NHT, without unitarity
constraints}

\end{figure}

Figure~\ref{fig-nht-no-ut} shows the pull distribution for the NHT
approach without unitarity constraints. The NHT approach produced
unbiased estimates of \(|V_{cb}|\), but showed some under-coverage
issues, indicating that the uncertainty might be underestimated. As
expected, the NHT tended to select simpler models (e.g., order (2,1,1)).
Without unitarity constraints, some fits saturated the unitarity bounds.

\begin{figure}

\centering{

\includegraphics[width=0.75\textwidth,height=\textheight]{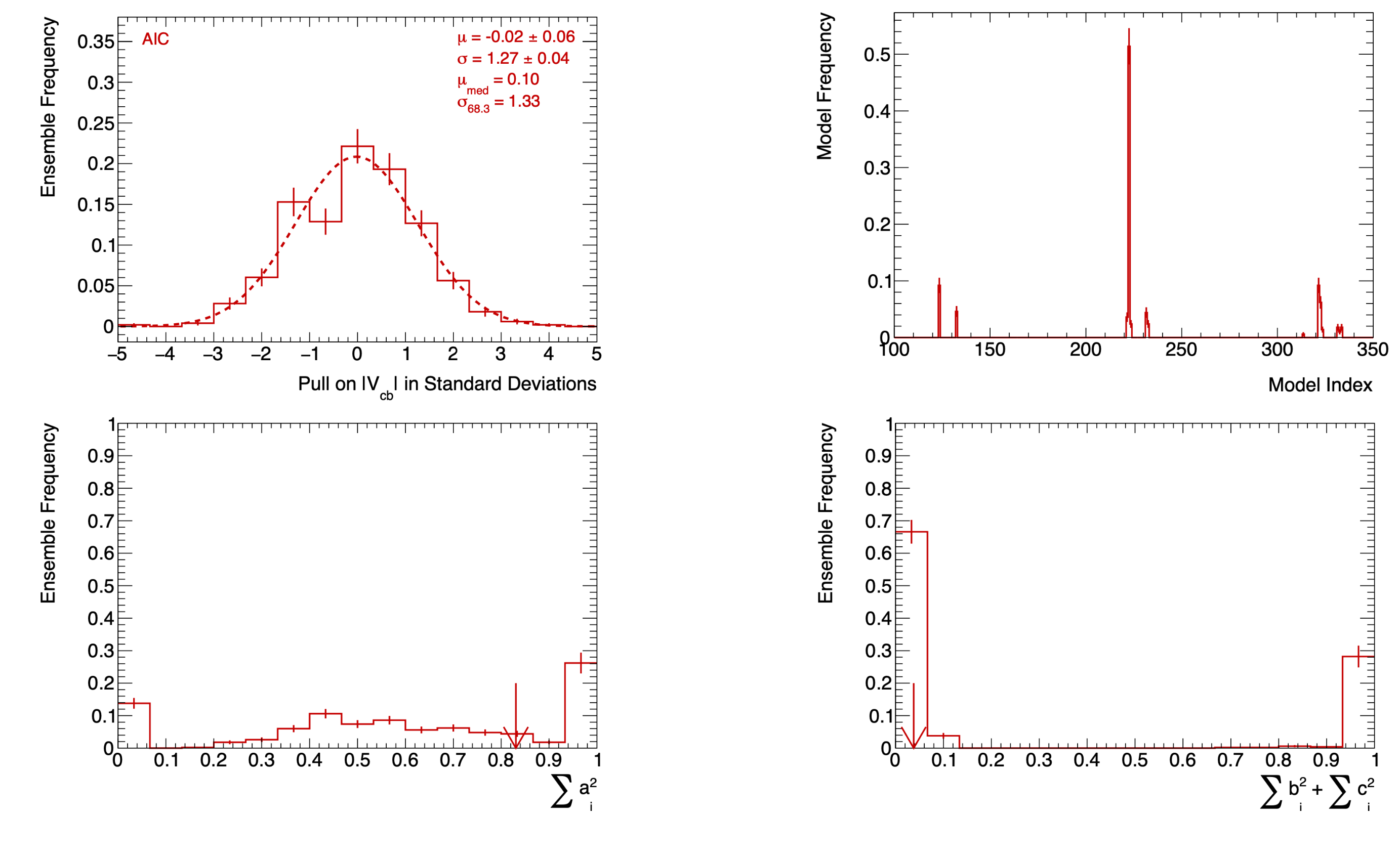}

}

\caption{\label{fig-aic-no-ut}Pull plot of AIC, no unitarity
constraints}

\end{figure}

Figure~\ref{fig-aic-no-ut} presents the pull distribution for the AIC
approach without unitarity constraints. The AIC approach showed
competitive performance compared to NHT; it produced similarly unbiased
estimates of \(|V_{cb}|\), but suffered from under-coverage issues
similar to NHT.

These results suggest that our AIC-based approach is a viable
alternative to the NHT method, offering comparable performance while
providing a simpler and statistically more rigorous framework for model
selection.

\subsection{Unitarity Constraints}\label{unitarity-constraints}

In addition to comparing the NHT and AIC approaches, we investigated the
impact of imposing unitarity constraints on our model selection
procedures.

\begin{figure}

\centering{

\includegraphics[width=0.75\textwidth,height=\textheight]{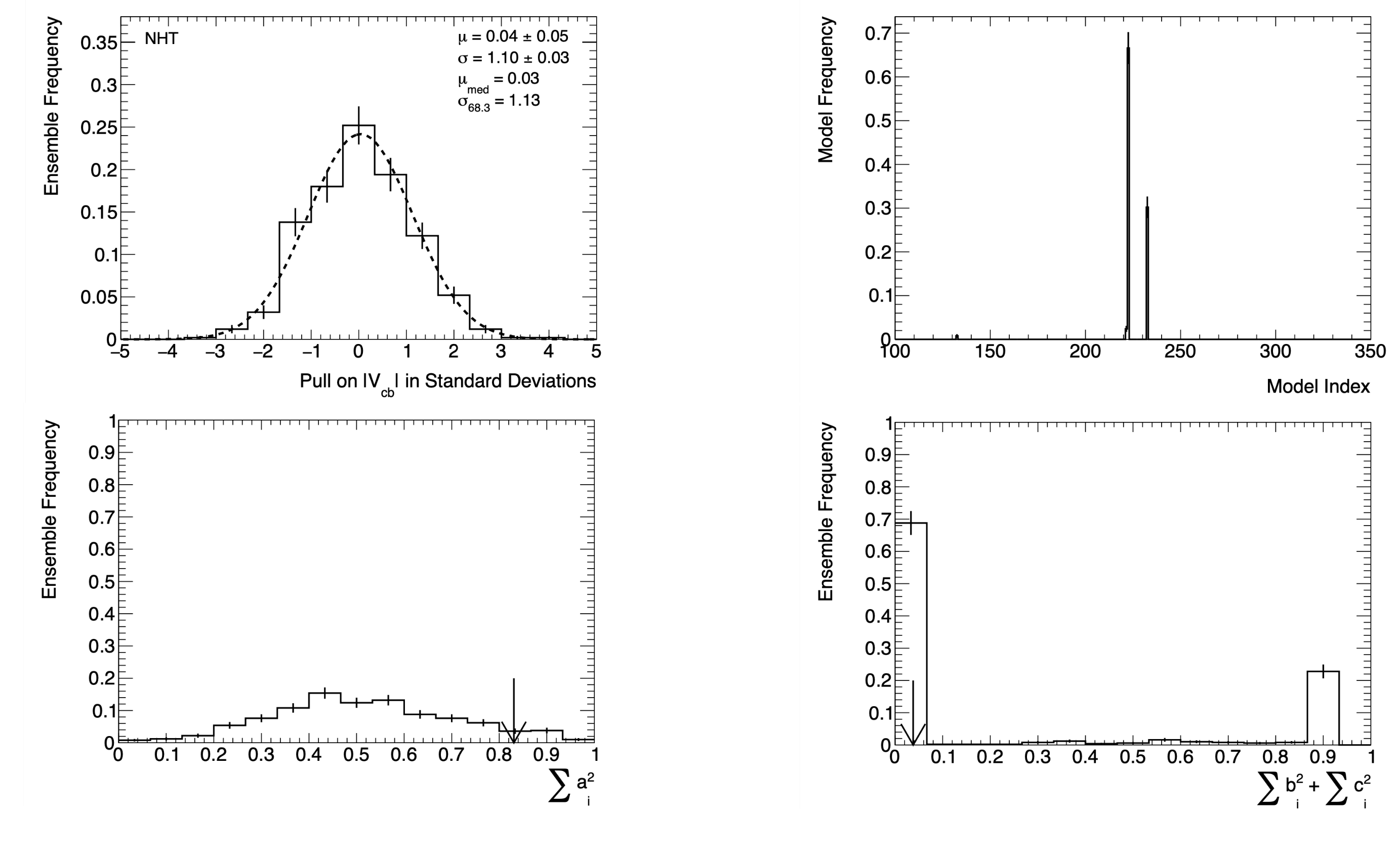}

}

\caption{\label{fig-nht-ut}Pull plot of NHT, with unitarity constraints}

\end{figure}

\begin{figure}

\centering{

\includegraphics[width=0.75\textwidth,height=\textheight]{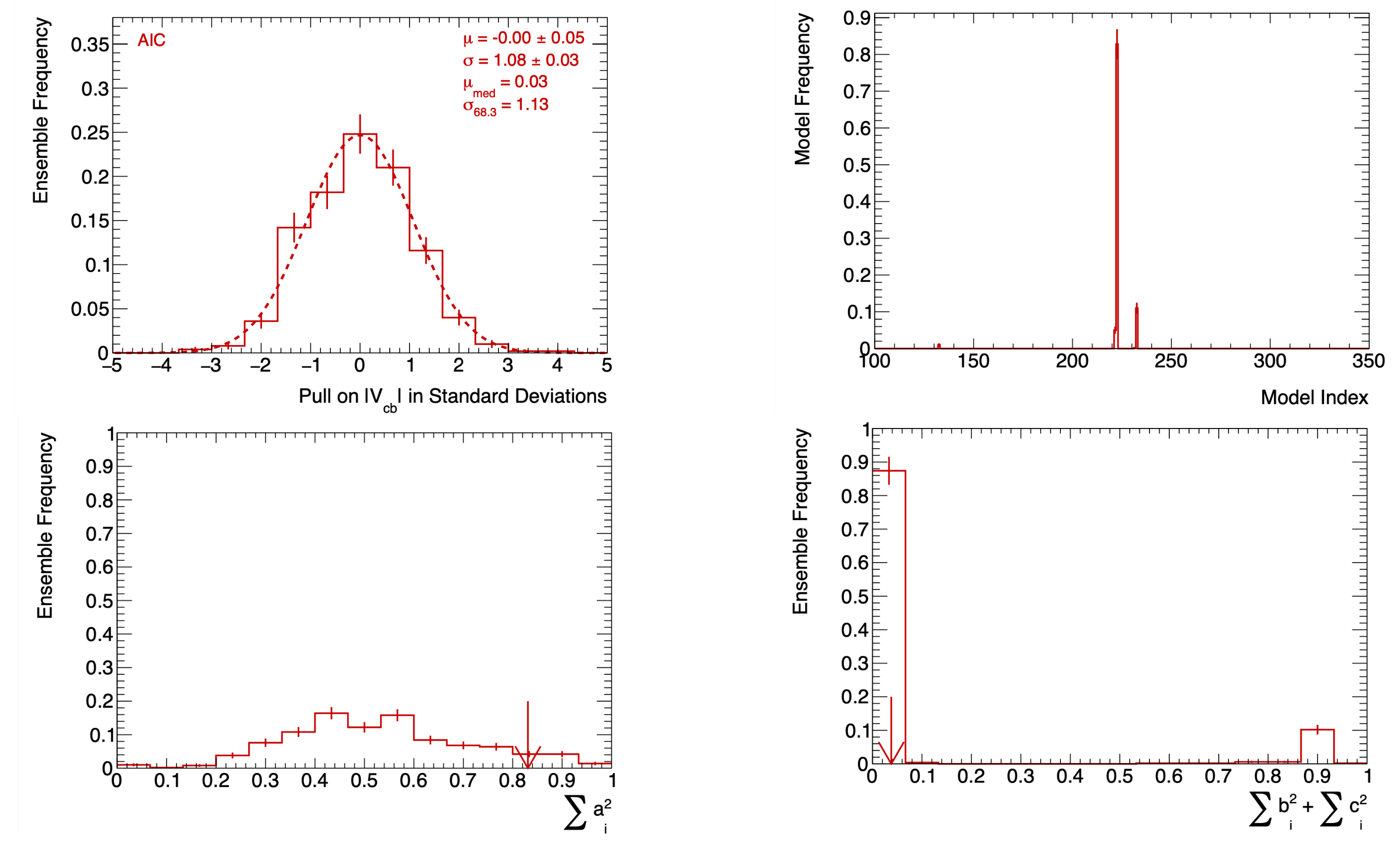}

}

\caption{\label{fig-aic-ut}Pull plot of AIC, with unitarity constraints}

\end{figure}

For both NHT and AIC approaches, imposing unitarity constraints
ameliorated the under-coverage issues observed in the unconstrained
fits. Figure~\ref{fig-nht-ut} shows the pull distribution for the NHT
approach with unitarity constraints, while Figure~\ref{fig-aic-ut}
presents the corresponding results for the AIC approach.

The relationship between unitarity constraints and coverage properties
is a key area for future investigation (F. Bernlochner et al.
forthcoming). Our results currently suggest that imposing unitarity
constraints can improve the coverage properties of \(|V_{cb}|\)
estimates, but the underlying reasons for this improvement remain
unclear.

\subsection{Global AIC: Model
Averaging}\label{global-aic-model-averaging}

Moving beyond single model selection, we also explored a model averaging
approach using the concept of Global AIC (gAIC) (Burnham and Anderson
1998). The key idea of gAIC is to weigh multiple models based on their
relative support from the data, rather than selecting a single ``best''
model. This approach can be formalized as follows:

\begin{enumerate}
\def\labelenumi{\arabic{enumi}.}
\tightlist
\item
  Calculate the AIC for each model.
\item
  Compute model weights based on the relative likelihood:
\end{enumerate}

\begin{equation}\phantomsection\label{eq-gaic-weights}{
w_i = \exp ( - \frac 1 2 \Delta_i )  / \sum_j \exp ( - \frac 1 2 \Delta_j )
}\end{equation}

where \begin{equation}\phantomsection\label{eq-gaic-delta}{
\Delta_i  = \text{AIC}_i - \text{AIC}_{\mathrm{min}}
}\end{equation}

\begin{enumerate}
\def\labelenumi{\arabic{enumi}.}
\setcounter{enumi}{2}
\tightlist
\item
  Calculate a weighted average of \(|V_{cb}|\) estimates across models:
\end{enumerate}

\begin{equation}\phantomsection\label{eq-gaic-vcb}{
V_{cb}| = \sum_i w_i |V_{cb}|_i
}\end{equation}

The estimator for the variance of the gAIC is given by:

\begin{equation}\phantomsection\label{eq-gaic-variance}{
\widehat{\text{var}(\hat{\theta})} = \left[\sum_{i=1}^R w_i \sqrt{\widehat{\text{var}(\hat{\theta}_i|g_i)} + (\hat{\theta}_i - \hat{\theta})^2}\right]^2
}\end{equation}

where \(\hat{\bar{\theta}} = \sum_{i=1}^R w_i \hat{\theta}_i\).

The gAIC approach offers several potential advantages. By considering
multiple models, it can better account for model uncertainty,
potentially leading to more robust estimates of \(|V_{cb}|\). It also
provides a natural way to incorporate information from all plausible
models, rather than relying on a single selected model which may not be
definitively better than its close competitors.

\begin{figure}

\centering{

\includegraphics[width=0.75\textwidth,height=\textheight]{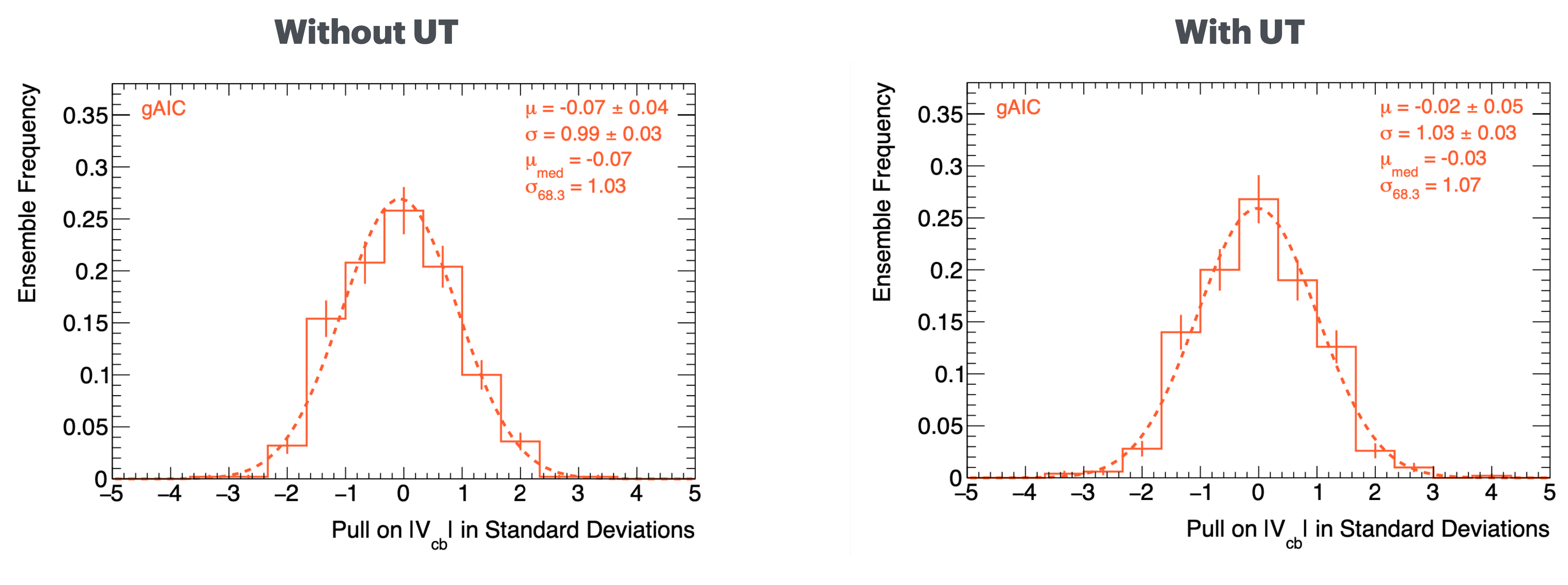}

}

\caption{\label{fig-gaic}Pull plot of gAIC, with and without unitarity
constraints}

\end{figure}

Figure~\ref{fig-gaic} shows the pull distribution for the gAIC approach,
both with and without unitarity constraints. The results from the gAIC
approach are particularly encouraging, since it produced unbiased
estimates of \(|V_{cb}|\) with correct coverage properties, both with
and without unitarity constraints.

The superior performance of gAIC, particularly in terms of coverage
properties, suggests that model averaging may be a valuable tool for
future \(|V_{cb}|\) determinations.

\section{Discussion and Future Work}\label{discussion-and-future-work}

Our results demonstrate the potential of rigorous model selection
techniques, particularly those based on information criteria and model
averaging, for improving \(|V_{cb}|\) extraction from exclusive
\(B \to D^* l \nu\) decays. The AIC-based approach, especially when
combined with model averaging in the form of gAIC, shows promise in
producing unbiased estimates of \(|V_{cb}|\) with correct coverage
properties.

However, several areas require further investigation:

\begin{enumerate}
\def\labelenumi{\arabic{enumi}.}
\tightlist
\item
  The source of under-coverage in non-averaged approaches, particularly
  when unitarity constraints are not imposed, needs to be better
  understood.
\item
  While our results support the use of AIC, it would be valuable to
  evaluate other model selection metrics to confirm AIC's superiority in
  this context. Potential alternatives include the Bayesian Information
  Criterion (BIC) (Schwarz 1978) or cross-validation approaches.
\item
  Incorporating external constraints, such as lattice QCD results, into
  our model selection framework presents both an opportunity and a
  challenge. Recent lattice QCD calculations (Bazavov et al. 2022;
  Harrison and Davies 2024; Aoki et al. 2024) provide valuable
  information about the form factors, but integrating this information
  into our model selection procedure requires careful consideration.
\end{enumerate}

\section{Conclusion}\label{conclusion}

In this work, we have presented a comprehensive study of model selection
techniques for \(|V_{cb}|\) extraction from exclusive
\(B \to D^* l \nu\) decays. We have demonstrated that an approach based
on the Akaike Information Criterion, particularly when combined with
model averaging, can produce unbiased estimates of \(|V_{cb}|\) with
correct coverage properties.

Our results highlight the importance of rigorous model selection in
precision flavor physics. By providing a principled way to choose
between different truncations of the BGL expansion, our approach reduces
the impact of arbitrary choices on \(|V_{cb}|\) determinations. The use
of model averaging further enhances the robustness of our results by
accounting for model uncertainty.

\section*{Bibliography}\label{bibliography}
\addcontentsline{toc}{section}{Bibliography}

\phantomsection\label{refs}

\end{document}